\documentclass{icrc}
            \usepackage{times}
\usepackage{graphicx}   
\begin{document}
\hyphenation{flu-xes}
\hyphenation{so-ur-ce}
\hyphenation{ba-ck-gro-und}
\hyphenation{mu-ons}
\hyphenation{co-m-pu-ta-ti-on}
\hyphenation{pa-ra-me-te-ri-z-a-t-i-on}
\hyphenation{pro-g-r-ess}
\hyphenation{se-c-t-i-ons}
\hyphenation{bre-m-s-s-t-r-a-h-l-ung}
\hyphenation{te-le-s-co-p-es}
\hyphenation{So-k-a-l-s-ki}
\hyphenation{Spi-e-r-ing}
\hyphenation{an-a-l-y-t-i-c-al}
\hyphenation{me-th-ods}
\hyphenation{us-es}
\hyphenation{Mo-n-te}
\hyphenation{Ca-r-lo}
\hyphenation{te-ch-n-i-que}
\hyphenation{ex-p-e-c-t-ed}
\hyphenation{de-t-e-c-t-or}
\hyphenation{re-s-p-o-n-se}
\hyphenation{mu-on}
\hyphenation{tra-n-s-p-o-r-t-a-t-i-on}
\hyphenation{abo-ve}
\hyphenation{un-c-e-r-t-a-i-n-t-i-es}
\hyphenation{se-c-t-i-on}
\hyphenation{op-t-i-m-i-s-t-ic}
\hyphenation{ev-a-l-u-a-t-i-on}
\hyphenation{su-r-v-i-v-al}
\hyphenation{pro-b-a-b-i-l-i-t-i-es}
\hyphenation{co-m-p-u-t-ed}
\hyphenation{PRO-P-MU}
\hyphenation{MU-S-IC}
\hyphenation{si-m-u-l-a-t-i-ons}
\hyphenation{gi-ve}
\hyphenation{pra-c-t-i-c-a-l-ly}
\hyphenation{sa-me}

\title{MUM: flexible precise algorithm for the muon propagation}
\author[1,2]{I. Sokalski}
\affil[1]{Institute for Nuclear Research, Russian Academy of Science, 
          Moscow 117312, Russia}
\affil[2]{now at DAPNIA/SPP, CEA/Saclay, 91191 Gif-sur-Yvette CEDEX, France}
\author[1]{E. Bugaev}
\author[1]{S. Klimushin}
\correspondence{I. Sokalski (sokalski@hep.saclay.cea.fr)}
\firstpage{1}
\pubyear{2001}
\maketitle

\begin{abstract}
We present a new muon propagation Monte Carlo FORTRAN code MUM (MUons+Medium) 
which possesses some advantages over analogous codes presently in use. The 
most important features of the algorithm are described. Data on the test for 
algorithm accuracy are presented. Contributions of different sources to the 
resulting error of simulation are considered. Selected results obtained with 
MUM are given and compared with ones from other codes.
\end{abstract}

\section{Introduction}
Propagation of muons in medium plays an important role for underground 
(-water, -ice) experiments with natural fluxes of high energy (HE) neutrinos 
and muons. Firstly, neutrinos are detected by muons which are born in $\nu N$ 
interactions and propagate a distance in medium from the point of interaction 
to a detector. Secondly, muons which are produced in atmospheric showers 
generated by cosmic rays represent the principal background for neutrino 
signal and therefore their flux at large depths should be well known. Besides,
the atmospheric muons deep under sea or earth surface are the only natural 
calibration source which allows to confirm correctness of the detector model by
comparison experimental and expected detector response. For the muon 
propagation along with analytical methods one uses the Monte Carlo (MC) 
technique which directly accounts for stochastic nature of the muon energy 
losses. There are several MC muon transportation algorithms currently in use 
(see, e.g. review in \citep{rhode1}) but theoretical and experimental progress
makes to create new ones. 

Here we present a new MC muon propagation code MUM (MUons+Medium) written in 
FORTRAN. When working on MUM we aimed at creation of an algorithm which would: 
{\it (a)} account for the most recent corrections for the muon cross-sections;
{\it (b)} be of adequate and known accuracy, i.e., does not contribute an 
additional systematic error which would exceed one from ``insuperable'' 
uncertainties (i.e. muon and neutrino spectra and cross-sections);
{\it (c)} be flexible enough, i.e. could be easily optimized for concrete 
purpose to desirable and well understood equilibrium between CPU time and 
accuracy and easily extended for any medium and any correction for the
cross-sections of the processes in which HE muon looses its energy;
{\it (d)} be ``transparent'', i.e. provide user with the whole set of data
related to used models for the muon cross-sections, energy losses, etc.;    
{\it (e)} be as fast as possible.
The MUM code has been developed for the Baikal experiment \citep{NT1,NT2} but
we believe it to be useful also for other experiments with natural fluxes of 
HE muons and neutrinos. 

\section{The basic features of the MUM algorithm~\footnote{
Detailed description can be found in \citep{MUMpres}}}

\begin{figure}[t]
\includegraphics[width=8.3cm]{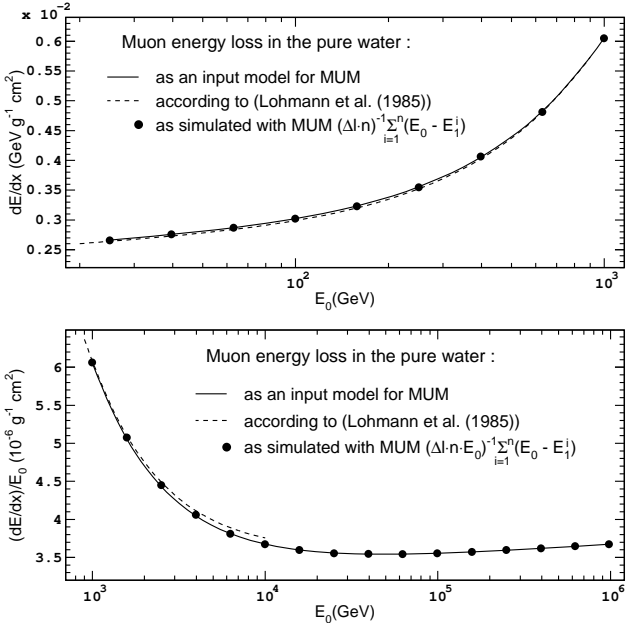}
\protect\caption{
The {\it simulated} with MUM muon energy losses (markers) and {\it model of 
energy losses} as used by MUM (solid lines). Also results on the muon energy 
losses from \protect\citep{lohman} are presented (dashed lines). The plot 
corresponds to simulation in pure water with $v_{cut} =$ 10$^{-2}$. 
}
\label{test1}
\end{figure}

\begin{figure}[t]
\includegraphics[width=8.3cm]{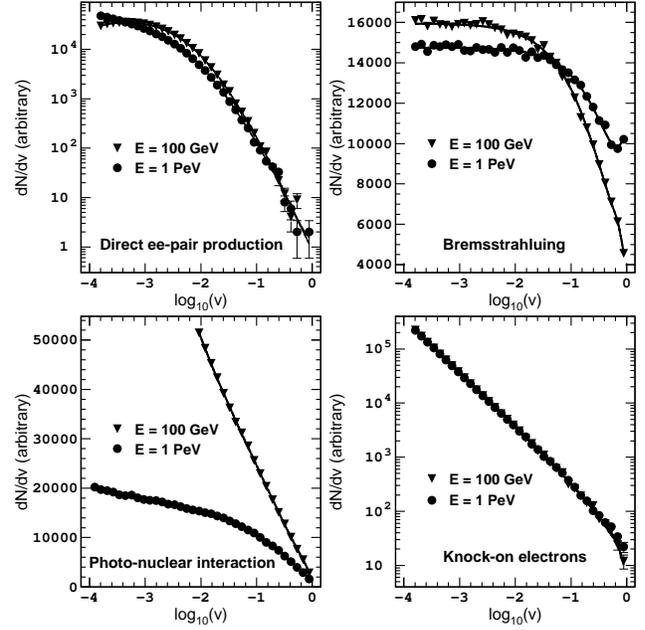}
\protect\caption{
Simulated with MUM distributions for the fraction of energy lost in a single 
interaction $v$ for muons with $E =$ 100 GeV (triangles) and $E =$ 1 PeV 
(circles) in comparison with corresponding differential cross-sections 
(lines). The case for pure water and $v_{cut} =$ 10$^{-4}$ is presented. The 
scales on Y-axis for $E =$ 100 GeV and $E =$ 1 PeV are different.
}
\label{test2}
\end{figure}

To get finite CPU time $T_{CPU}$, the energy losses in any MC muon propagation
algorithm have to be decomposed into two parts: muon interactions with 
fraction of energy lost $v$ which exceeds some value $v_{cut}$, are simulated 
directly while the part of interaction with $v < v_{cut}$ is treated by the 
approximate concept of ``continuum'' energy loss. Setting $v_{cut}$ too low 
one looses the speed (roughly, $T_{CPU} \propto v^{-1}_{cut}$) but setting it 
too high, one looses the accuracy. We did not fix $v_{cut}$ in MUM, it may be 
set optionally within a range of 10$^{-4} \le v_{cut} \le$ 0.2, since the 
optimum value depends on the concrete case \citep{MUMer}. 

An ``absolute'' energy transfer threshold $\Delta E_{cut}$ in a range of 10 
MeV $\le \Delta E_{cut} \le$ 500 MeV can be used in MUM along with 
``relative'' threshold $v_{cut}$ to simulate the muon interactions within 
detector sensitive volume. It is important for deep underwater (-ice) 
Cherenkov neutrino telescopes 
\citep{NT1,NT2,AMANDA1,AMANDA2,ANTARES1,ANTARES2,NESTOR1,NESTOR2}, where the 
water or ice are used both as a shield which absorbs atmospheric muons and as
a detecting medium.

The formulae for the cross-sections of muon interactions (bremsstrahlung,
$e^{+}e^{-}$-pair production, photo-nuclear interaction, knock-on electron
production) are given in \citep{MUMpres}. The code does not use any 
preliminary computed files, all necessary data are prepared at the initiation 
on the base of several relatively short routines, which give cross-sections 
for the muon interactions and stopping-power formula for ionization. It allows
user to correct or even entirely change the model for any type of the muon 
interaction.

At each step we tried to avoid when possible any simplifications when 
computing/simulating the free path between two interactions, energy transfers,
etc., or, at least, to track the error which comes from this or that kind of 
simplification. In most cases, to keep $T_{CPU}$ at low level, roots for 
equations and values for functions are found at initiation, tabulated and then
referenced when necessary by a interpolation algorithm which was carefully 
checked for each case to guarantee the high enough level of accuracy.

Formally, MUM simulates the propagation of the muons with the energies up to 1
EeV but one should keep in mind that above 1 PeV uncertainties with muon 
cross-sections grow remarkably, some effects which expose at UHE (e.g., LPM 
effect) are not accounted in MUM. Three media are available instantly for the 
muon propagation with MUM, namely pure water, ice and standard rock. But any 
medium can be easily composed by user following examples which are given in 
the initiation routine. At its current version MUM represents an 1D--algorithm
which does not track the angular and lateral deviations of muons, but it is 
planned to be 3D--extended.

\section{The accuracy of the algorithm and an optimum setting of simulation 
         parameters}

\begin{figure}[t]
\includegraphics[width=8.3cm]{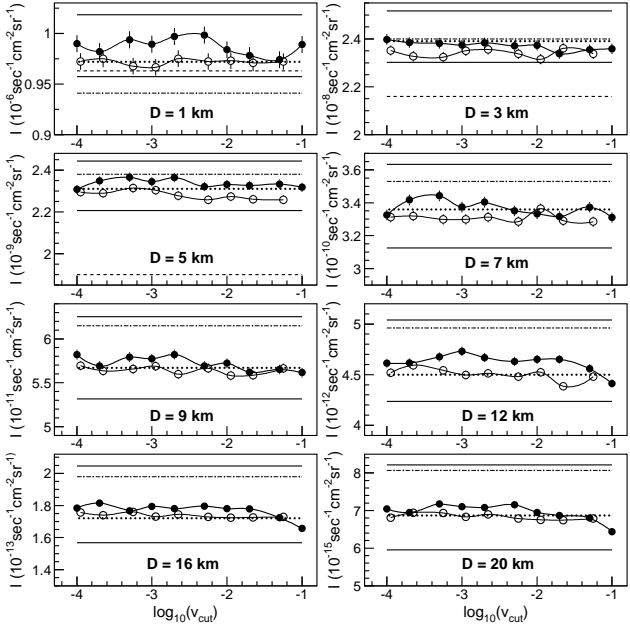}
\protect\caption{
Intensity of vertical atmospheric muon flux $I$ at different depths $D$ of 
pure water vs. $v_{cut}$ as obtained by simulation with MUM. Muons were 
sampled according to sea-level spectrum from \protect\citep{bks1}. Closed 
circles: knock-on electron production with fraction of energy lost 
$v \ge v_{cut}$ is simulated; open circles: ionization is completely 
``continuous''. Two horizontal solid lines on each plot show the flux 
intensity simulated with all muon cross-sections multiplied by a factor 1.01 
(lower line) and 0.99 (upper line) for $v_{cut}$ = 10$^{-4}$. Dashed lines on 
plots for $D \le$ 5 km correspond to intensity which was calculated for all 
energy loss treated as ``continuous''. Dash-dotted lines show the flux 
intensity simulated with muons sampled according to sea level spectrum 
\protect\citep{gaisser} and $v_{cut}$ = 10$^{-4}$. Dotted lines 
correspond to $v_{cut}$ = 10$^{-4}$ and cross-section for absorption of a real
photon at photo-nuclear interaction parameterized according to 
\protect\citep{ZEUS} instead of parameterization \protect\citep{phnubb} which 
is the basic in MUM.
}
\label{test3}
\end{figure}

Fig.1 shows results of an accuracy test which consisted of following. For a 
muon energy $E_{0}$ the short distance $\Delta l$ was chosen and $n$ muons 
were propagated through this distance with MUM~\footnote{The value 
$l = n \cdot \Delta l$ must be much greater than mean free path $\bar L(E_0)$ 
between two interactions with $v > v_{cut}$ to obtain statistically significant
result but also must be small enough since there should be no stopped muons.}. 
Then, $(E_0 - \frac{1}{n}\;\sum^{n}_{i=1}\;E^{i}_1)/\Delta l$ (where 
$E^{i}_{1}$ is energy of $i$-th muon at the end of distance $\Delta l$) 
represents energy losses {\it simulated} by the algorithm. In an ideal case 
they should be equal to ones which can be directly calculated by integration 
of differential cross-sections which represents the {\it incoming model} for 
given code but, since algorithm itself necessarily contributes an error which 
originates from application of numerical procedures which the code consists 
of, the simulated energy losses and incoming model for energy 
losses are not the same for any real code. The only case is presented in Fig.1
but actually such test was performed both for water and standard rock with 
$v_{cut}$ in a range from 10$^{-4}$ to 0.05 \citep{MUMpres}. The difference 
between simulated energy losses and incoming model for energy losses for MUM 
does not exceed 1\% 
except for the case when $v_{cut} \ge$ 0.01 and ionization is treated as 
completely ``continuous'' process. It means that inner inaccuracy of 
MUM contributes to the resulting error much less than principal uncertainties 
with muons and neutrinos fluxes, cross-sections, etc. 
 
Fig.2 demonstrates the accuracy of simulation for fraction of energy lost $v$ 
for different kinds of muon interaction. Simulated distributions are plotted 
along with functions for differential cross-sections $d\sigma / dv$. Again, 
only small part of the data is shown in the plot but agreement between 
simulated distributions and predictions is not worse for other media and other
muon energies which has been tested carefully. 

Results on simulation of atmospheric muon vertical flux at different depths in 
the pure water vs. $v_{cut}$ are shown in Fig.3. Simulations were performed 
for 2 atmospheric muon surface spectra; with knock-on electron production
included in simulation of energy losses and treated as totally ``continuous'';
with 2 different parameterizations for photo-nuclear interaction; with all 
muon energy losses multiplied by 0.99, 1.00 and 1.01 (which corresponds to 
the most optimistic evaluation for uncertainties with the muon cross-sections 
of 1\% 
\citep{rhode1,kp}). The general conclusion is as follows: the principal 
uncertainties when computing the atmospheric muon flux at large depth are ones
for the muon cross-sections. Influence of set value for $v_{cut}$ in a range 
10$^{-4}$--10$^{-1}$ is much less and, in principal, in a ``ideal muon 
propagation code'' one could set $v_{cut} =$ 10$^{-1}$ which allows 
calculations to be rather fast without remarkable influence on the result. 
Also the ionization energy losses can be treated as completely ``continuous'' 
(which saves $T_{CPU}$ with a factor of $\sim$2). But, as MUM's own accuracy 
(in the sense of reproducing the muon energy losses) becomes worse than 1\% 
for $v_{cut} \ge$ 10$^{-2}$ if ionization is excluded out of simulation and 
for $v_{cut} \ge$ 5$\cdot$10$^{-2}$ if knock-on electrons are simulated
\citep{MUMpres} we conservatively affirm $v_{cut} =$ 5$\cdot$10$^{-2}$ and
knock-on electron production included in simulation as a optimum setting of 
parameters for simulation the atmospheric muon flux at large depths  with MUM. 
With such setting the proportion of $T_{CPU}$ which is necessary to get the 
same statistics with muon propagation algorithms MUM, PROPMU \citep{lipari} and
MUSIC \citep{music1} is approximately $1 : 10 : 600$ (note that MUM, in 
contrast both to PROPMU and MUSIC, is 1D algorithm). 

We did not investigate 
specially the influence of simulation parameters on the results for the muon 
flux originated from neutrino. Generally, intensity of the muon flux 
$I^{AC}_{\mu}$ which accompanies the neutrino flux in a medium is proportional
to the muon range, and, consequently, $I^{AC}_{\mu} \propto (dE/dx)^{-1}$ (in 
contrast to atmospheric muons whose flux at the large depths depends more 
sharply upon muon energy losses - see, e.g., Fig.3). 
That means that an error for simulated flux of muons produced by neutrino 
is proportional to an error in muon energy losses.
So, the setting of 
parameters described above fits even better for propagation of muons 
originated from neutrino. 

\section{Comparison to other muon propagation algorithms}

\begin{figure}[t]
\includegraphics[width=8.3cm]{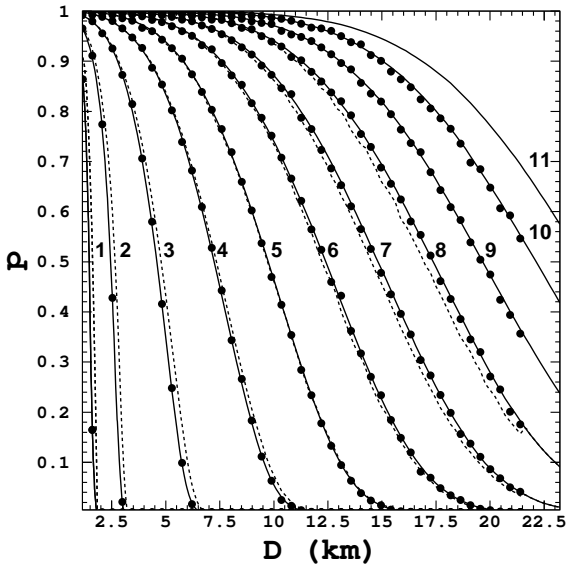}
\protect\caption{
Survival probabilities vs. distance of propagation in pure water as simulated
with MUM (solid lines), PROPMU (dashed lines) and MUSIC (circles). Figures 
near curves indicate initial energy of mono-energetic muon beam as follows: 
500 GeV (1), 1 TeV (2), 3 TeV (3), 10 TeV (4), 30 TeV (5), 100 TeV (6), 
300 TeV (7), 1 PeV (8), 3 PeV (9), 10 PeV (10), 30 PeV (11).
}
\label{sp}
\end{figure}

\begin{figure}[t]
\includegraphics[width=8.3cm]{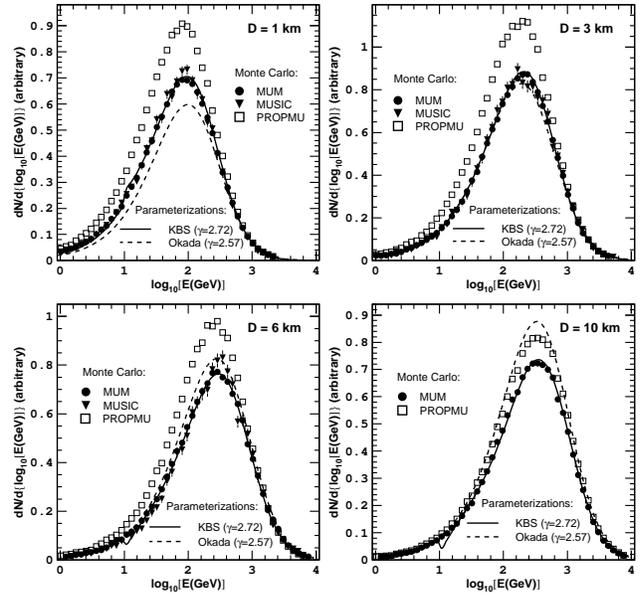}
\protect\caption{
Differential spectra of vertical atmospheric muons at four depths in the pure
water as simulated with MUM, PROPMU and MUSIC and parameterized by 
\protect\citep{bks1,okada}.   
}
\label{sp1}
\end{figure}

Fig.~\ref{sp} shows  survival probabilities (fractions of muons which have 
survived after propagation of distance $D$) vs. distance of propagation in 
pure water as computed with MUM, PROPMU and MUSIC for a set of muon energies 
from 500 GeV to 30 PeV. There are no statistically significant differences 
between MUM and MUSIC but survival probabilities computed with PROPMU are 
noticeably higher at energies $E \le$ 30 TeV (up to a factor of 6) and become 
less at $E \ge$ 30 TeV.

In Fig.~\ref{sp1} differential spectra for vertical atmospheric muons at 
different depths in pure water are presented as simulated with MUM, PROPMU and
MUSIC. Muons at the surface were sampled according to spectrum \citep{bks1}. 
Okada \citep{okada} and KBS \citep{bks1} parameterizations for deep underwater
muon spectra are shown, as well. MUSIC and MUM give almost the same results 
because survival probabilities for muons in pure water are the same when 
simulating with MUSIC and MUM. MUSIC's and MUM's spectra coincide with KBS 
parameterization which is based on the same sea-level muon spectrum as was 
used for simulation and on muon propagation with MUM. Okada parameterization is
lower than KBS, MUM and MUSIC results (up to 18\% 
in terms of integral muon flux at $D =$ 1 km) at relatively shallow depths and
becomes higher at $D \ge$ 5 km because it is based on rather hard surface 
muon spectrum with index $\gamma =$ 2.57 which leads to a deficit for low 
energy muons comparing to KBS parameterization. Simulation with PROPMU 
produces the muon spectra which {\it i)} are significantly higher 
(31\%, 30\%, 27\% and 17\% 
in terms of integral muon flux at the depths $D =$ 1 km, 3 km, 6 km  and 10 
km, correspondingly) and {\it ii)} are expanded to the low energies. It is in 
qualitative agreement with results on survival probabilities presented in 
Fig.~\ref{sp}. 

\section{Conclusions}

We have presented the muon transportation algorithm MUM and have given 
selected results obtained with it in comparison with ones obtained with 
analogous codes. We consider the current version of MUM as a basis for the 
further development. The code is available by request. 

\begin{acknowledgements}
We are grateful to I. Belolaptikov, A. Butkevich, R. Kokoulin, V. Kudryavzev, 
P. Lipari, W. Lohmann, V. Naumov, O. Streicher and Ch. Wiebusch for their 
comments and critic. 
\end{acknowledgements}

\end{document}